\documentclass[prb,twocolumn,aps]{revtex4}
\usepackage{graphicx}
\usepackage{dcolumn}
\usepackage{bm}
\usepackage{hyperref}
\renewcommand{\vec}[1]{{\mathbf{#1}}}
\newcommand{\beq}{\begin{eqnarray}}
\newcommand{\eeq}{\end{eqnarray}}
\newcommand{\myfig}[3]

\begin{document}
\title{Hidden Charge 2e Boson: Experimental Consequences for Doped Mott Insulators}
\author{Ting-Pong Choy$^{a,b}$}
\author{Robert G. Leigh$^a$}
\author{Philip Phillips$^{a,b}$}
\affiliation{$^a$Department of Physics,
University of Illinois
1110 W. Green Street, Urbana, IL 61801, U.S.A.}
\affiliation{$^b$Kavli Institute for Theoretical Physics, University
of California, Santa Barbara, CA 93106-4030}
\date{\today}

\begin{abstract}
We show here that many of the normal state properties of the
cuprates can result from the new charge 2e bosonic field which we
have recently (Phys. Rev. Lett. {\bf 99}, 46404 (2007)  and Phys.
Rev. B 77, 014512 (2008)) shown to exist in the exact low-energy
theory of a doped Mott insulator. In particular, the 1) mid-infrared
band including the non-vanishing of the restricted f-sum rule in the
Mott insulator, 2) the $T^2$ contribution to the thermal
conductivity, 3) the pseudogap, 4) the bifurcation of the electron
spectrum below the chemical potential as recently seen in
angle-resolved photoemission, 5) insulating behaviour away from
half-filling, 6) the high and low-energy kinks in the electron
dispersion and 7) T-linear resistivity all derive from the charge 2e
bosonic field.

We also calculate the inverse dielectric function and show that it
possesses a sharp quasiparticle peak and a broad particle-hole
continuum.  The sharp peak is mediated by a new charge e composite
excitation formed from the binding of a charge 2e boson and a hole
and represents a distinctly new prediction of this theory.  It is
this feature that is responsible for dynamical part of the spectral
weight transferred across the Mott gap.  We propose that electron
energy loss spectroscopy at finite momentum and frequency can be
used to probe the existence of such a sharp feature.
\end{abstract}

\maketitle

Key challenges facing a theory of the normal state (at zero magnetic
field) of the copper oxide high temperature superconductors include
1) T-linear resistivity\cite{batlogg}, 2) the
pseudogap\cite{alloul,norman,timusk}, 3) absence of
quasiparticles\cite{Kanigel}, and 4) the mid-infrared band in the
optical conductivity\cite{cooper,uchida1,opt0,opt1,opt2,opt3}. Since
the parent cuprates are Mott insulators, the normal state properties
should, in principle, be derivable from the corresponding low-energy
theory.   Proper low-energy theories are constructed by integrating
out the degrees of freedom far away from the chemical potential.

While no shortage of low-energy theories has been
proposed\cite{sl1,sl2,sl3,sl4,sl5,sl6,sl7,sl8,sl9,pert0,pert1,pert2,pert3,pert33,pert4,pert5,slave1,slave2,slave3},
none, until recently\cite{ftm4prl,ftm4prb}, has been based on an
explicit integration of the degrees of freedom at high energy, even
in the simplest model for a doped Mott insulator,

\beq
H_{\rm Hubb}&=&-t\sum_{i,j,\sigma} g_{ij} c^\dagger_{i,\sigma}c_{j,\sigma}+U\sum_{i,\sigma} c^\dagger_{i,\uparrow}c^\dagger_{i,\downarrow}c_{i,\downarrow}c_{i,\uparrow}.\nonumber\\
\eeq

Here $i,j$ label lattice sites, $g_{ij}$ is equal to one iff $i,j$
are nearest neighbours, $c_{i\sigma}$ annihilates an electron with
spin $\sigma$ on lattice site $i$, $t$ is the nearest-neighbour
hopping matrix element and $U$ the energy cost when two electrons
doubly occupy the same site. The cuprates live in the strongly
coupled regime in which the interactions dominate as $t\approx
0.5$eV and $U=4$eV. As $U$ is the largest energy scale, it is
appropriate to integrate over the fields that generate the physics
on the $U$ scale.  We showed\cite{ftm4prl,ftm4prb} explicitly how to
perform such an integration.  The corresponding low-energy theory
contains degrees of freedom, namely a charge 2e boson, that are 1)
not in the original model and 2) more importantly, are not made out
of the elemental excitations.  We show here that this new degree of
freedom mediates many of the anomalies in the normal state of the
cuprates as probed by electrical and thermal transport as well as
angular-resolved photoemission spectroscopy (ARPES).

\section{Overview of Low-Energy Theory}

In a series of papers\cite{ftm4prl,ftm4prb}, we showed how to coarse
grain the Hubbard model cleanly for $U\gg t$. We accomplished this
by extending the Hilbert space of the Hubbard model and associating
with the high energy scale a new fermionic oscillator which is
constrained.  The coupling constant for the fermionic oscillator is
$U$.  We showed that once the constraint is solved, we obtained
exactly the Hubbard model.  However, since the energy scale of the
new osciallator is $U$, the low-energy theory is obtained by
integrating over this field.  We showed
explicitly\cite{ftm4prl,ftm4prb} that the Lagrangian in the extended
Hilbert space is quadratic in the high-energy field and hence all
integrals can be performed exactly.  Let us define

\beq\label{eom} {\cal
M}_{ij}=\delta_{ij}-\frac{t}{(\omega+U)}g_{ij}\sum_\sigma
c_{j,\sigma}^\dagger c_{i,\sigma}
\eeq

and $b_{i}=\sum_{j}b_{ij}=\sum_{j\sigma} g_{ij}c_{j,\sigma}V_\sigma
c_{i,-\sigma}$ with $V_\uparrow=-V_\downarrow=1$. At zero frequency,
the exact low-energy Hamiltonian is

\beq H^{IR}_h = -t\sum_{i,j,\sigma}g_{ij}
\alpha_{ij\sigma}c^\dagger_{i,\sigma}c_{j,\sigma}+ H_{\rm
int}-\frac{1}{\beta}Tr\ln{\cal M}, \nonumber
\eeq

where

\beq\label{HIR}
H_{\rm int}=-\frac{t^2}U \sum_{j,k} b^\dagger_{j}
({\cal M}^{-1})_{jk} b_{k}-\frac{t^2}U\sum_{i,j}\varphi_i^\dagger
 ({\cal M}^{-1})_{ij} \varphi_j\nonumber\\
-t\sum_j\varphi_j^\dagger c_{j,\uparrow}c_{j,\downarrow}-\frac{t^2}U \sum_{i,j}\varphi^\dagger_i ({\cal M}^{-1})_{ij}
b_{j}+h.c.\;\;.
\eeq

which contains a charge 2e bosonic field, $\varphi_i$.

The low-energy theory acts only in the original Hilbert space of the
Hubbard model because we integrated out all operators which acted in
the extended space.  Consequently,
it is incorrect to interpret $\varphi_i$ as a canonical boson operator
with an associated Fock space.  Likewise, we should not immediately
conclude that $\varphi_i$ gives rise to a propagating charge 2e
bosonic mode, as it does not have canonical kinetics; at the earliest,
this could be generated at order $O(t^3/U^2)$ in perturbation
theory. Alternatively, we believe that $\varphi$ appears as a bound
degree of freedom.  The new composite charge e state is crucial to explaining the origin of
dynamical spectral weight transfer across the Mott gap. As a consequence, the conserved charge is no longer just the electron number but rather,
\beq
Q=\sum_{i\sigma}c_{i\sigma}^\dagger c_{i\sigma}+2\sum_i\varphi^\dagger_i\varphi_i.
\eeq

The terms containing $\varphi_i$ are absent from projective theories and represent the fact that consistent with the Hubbard model, the corresponding
low-energy theory is not diagonal in any sector containing a fixed number of
doubly occupied sites.  As we have shown elsewhere\cite{tjredux}, the
 presence of double occupancy in the low-energy theory presented above is
completely compatible with the standard derivation of the t-J model.
One simply has to remember that the operators appearing in the t-J
model are transformed\cite{eskes,pert1,gingras} fermions which are
related by a similarity transformation to  the bare electrons
appearing in the Hubbard model.  If the rotated fermions are
expressed\cite{tjredux} in terms of the original electron operators
in the Hubbard model, double occupancy of the bare fermions is
reinstated. In fact, the t-J model expressed\cite{tjredux} in terms
of the original bare electron operators contains processes which mix the doubly and singly
occupied sectors with a matrix element $t^2/U$

(\ref{HIR}). However, it is common to
interpret\cite{sl1,sl2,sl3,sl4,sl5,sl6,sl7,sl8,sl9,pert0,pert2,pert3,pert33,pert4,pert5,slave1,slave2,slave3}
the $t-J$ model as a model about physically doped holes in a Mott
insulator. Unless the operators in the $t-J$ model are transformed
back to the original basis, it is impossible to make any connection
between doped holes in the two theories.

\begin{figure}
\centering
\includegraphics[width=4.5cm]{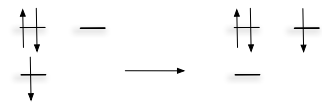}
\caption{Hopping processes mediated by the $\varphi_i$ operator in the
low-energy theory obtained by explicitly integrating out the high
energy sector.  This process is in the t-J model only if one tranforms
the electron operators back to the orignal Hubbard basis, retaining
the terms that change the number of doubly occupied sites.}
\label{midir}
\end{figure}

Physical processes mediated by the boson include singlet
motion (third term in Eq. (\ref{HIR})) and motion of double occupancy
in the lower Hubbard band (second term in Eq. (\ref{HIR})).  The latter has a bandwidth of $t$ and, as we will show, constitutes the mid-infrared band.  Since Eq. (\ref{HIR}) retains all the low-energy degrees of freedom, a $(t/U)$ expansion is warranted. To leading order in $t/U$, ${\cal M}=\delta_{ij}$ and the effective low-energy Hamiltonian simplifies to
\beq
\label{HIR-simp}
H_{\rm eff}&=&-t\sum_{i,j,\sigma}g_{ij} \alpha_{ij\sigma}c^\dagger_{i,\sigma}c_{j,\sigma}\nonumber\\
&&-\frac{t^2}U \sum_{j} b^\dagger_{j} b_{j}-\frac{t^2}U\sum_{i,j}\varphi_i^\dagger \varphi_i\nonumber\\
&&-t\sum_j\varphi_j^\dagger c_{j,\uparrow}c_{j,\downarrow}-\frac{t^2}U \sum_{i,j}\varphi^\dagger_i
b_{i}+h.c.\;\;.
\eeq
The first two terms contain the interactions in the $t-J$ model as the
two-site term in $b_j^\dagger b_j$ is proportional to the unprojected
spin-spin interaction.  In second-order perturbation theory, the
 interaction term $\varphi_i^\dagger b_i$ mediates the process in
 Fig. (\ref{midir}).   In this process, hole motion over three sites is possible
only if a doubly occupied site and a hole are neighbours.  This
process is absent if one assumes that the no-double occupancy condition
in the t-J model applies to the bare electrons as well.  The latter is
true exactly at $U=\infty$. As we will see, this
process, which is present as a result of the bosonic degree of freedom,
is responsible for many of the anomalous properties of the normal
state of the cuprates.

\subsection{Electron Spectral Function}

We analyze the nature of the electronic excitations at low energy by
focusing on the electron spectral function. Since this calculation
cannot be done exactly, we establish at the outset the basic physics
that a correct calculation should preserve. We have shown
previously\cite{ftm4prl,ftm4prb} that the electron creation operator
at low energy,
\beq\label{cop}
c^\dagger_{i,\sigma}&\rightarrow&(1-n_{i,-\sigma})c_{i,\sigma}^\dagger
+V_\sigma \frac{t}{U} b_i^\dagger{\cal M}_{ij}^{-1}
c_{j,-\sigma}\nonumber\\&-& V_\sigma
\frac{s}{U}\varphi_i^\dagger{\cal M}_{ij}^{-1}c_{j,-\sigma},
\eeq
contains the standard term for motion in the lower Hubbard band
(LHB), $(1-n_{i,-\sigma})c_{i,\sigma}^\dagger$
($n_{i,-\sigma}c_{i\sigma}$ in the upper Hubbard band (UHB) for
electron doping) with a renormalization from spin fluctuations
(second term) and a new charge $e$ excitation, $c_{i,-\sigma}{\cal
M}_{ij}^{-1}\varphi_j^\dagger$. In the lowest order in $t/U$, our
theory predicts that the new excitation corresponds to
$c_{i,-\sigma}\varphi_i^\dagger$, that is, a hole bound to the
charge $2e$ boson.  This extra charge e state mediates dynamical
(hopping-dependent) spectral weight transfer across the Mott gap.

Consequently, we predict that an electron at low energies is in a
superposition of the standard LHB state (modified with spin
fluctuations) and a new charge $e$ state which is a composite
excitation. It is the presence of these two distinct excitations
that gives rise to the static (state counting giving rise to
2x)\cite{sawatzky} and dynamical parts of the spectral weight
transfer. A saddle-point analysis will select a particular solution
in which $\varphi_i$ is non-zero.  This will not be consistent with
the general structure of Eq. (\ref{cop}) in which part of the
electronic states are not fixed by $\varphi_i$.  Similarly, mean
field theory in which $\varphi_i$ is assumed to condense, thereby
thwarting the possibility that new excitations form, is also
inadequate.

The procedure we adopt is the simplest that preserves the potential strong interactions between the Bose and Fermi degrees of freedom.  We treat $\varphi_i$ to be spatially independent with no dynamics of its own.  This interpretation is consistent with the fact that $\varphi_i$ acquires dynamics only through electron motion.  Under this assumption, the single-particle electron Green function
\beq
G(k,\omega)=-iFT \langle {Tc_i(t) c^\dagger_j(0)} \rangle,
\eeq
can be calculated rigorously in the path-integral formalism as
\begin{widetext}
\beq
G(k,\omega)= -i FT \int [D\varphi_i^*] [D\varphi_i] \int [Dc_i^*] [Dc_i]  c_i(t) c^*_j(0) \exp^{-\int L[c,\varphi] dt},
\eeq
\end{widetext}
where $FT$ refers to the Fourier transform and $\textbf T$ is the
time-ordering operation. Further, since our focus is on the new
physics mediated by the interaction between the Bose and Fermi
degrees of freedom and Eq. (\ref{cop}) suggests that the spin-spin
interaction simply renormalizes the electronic states in the LHB, we
will neglect the spin-spin term.  To see what purely fermionic model
underlies the neglect of the spin-spin term in Eq. (\ref{HIR-simp}),
we integrate over $\varphi_i$.  The full details of how to carry out
such an integration are detailed elsewhere\cite{ftm4prb}.  The resultant Hamiltonian is not the Hubbard model but rather,
\beq
H'=H_{Hubb}-\frac{t^2}{U}\sum_i b_i^\dagger b_i,
\eeq
a $t-J-U$ model in which the spin-exchange
interaction is not a free parameter but fixed to $J=-t^2/U$.
That the $t-J-U$ model with $J=-t^2/U$ is equivalent to a tractable IR model, namely
Eq. (\ref{HIR-simp}) without the spin-spin term, is an unexpected
simplification.  As Mott physics still pervades the $t-J-U$ model in the
vicinity of half-filling, our analysis should reveal the non-trivial
charge dynamics of this model. That is, the physics we uncloak here is
independent of the spin degrees of freedom.  In fact, what we show is
that the spin-spin interaction (as in the hard-projected $t-J$ model) is
at best ancillary to many of the normal state properties of the cuprates.

To proceed, we will organize the calculation of $G(k,\omega)$ by first integrating out the fermions (holding $\varphi$ fixed)
\begin{widetext}
\beq\label{geff}
G(k,\omega)= \int [D\varphi^*] [D\varphi] FT \left( {\int [Dc_i^*] [Dc_i]  c_i(t) c^*_j(0) \exp^{-\int L[c,\varphi] dt} }\right)
\eeq
\end{widetext}
where now
\begin{widetext}
\beq
L=\sum_{i\sigma}(1-n_{i\bar\sigma}) c_{k\sigma}^* \dot c_{k\sigma} -\left( { 2\mu + \frac{s^2}{U} }\right)\varphi^*\varphi -\sum_{k\sigma}(g_t t \alpha_k+\mu) c_{k\sigma}^* c_{k\sigma}+s\varphi^*\sum_k (1-\frac{2t}{U}) c_{-k\downarrow}c_{k\uparrow}+c.c.
\eeq
\end{widetext}
The effective Lagrangian can be diagonalized and written as
\begin{widetext}
\beq
\label{leff}
L=\sum_{k\sigma}(1-n_{i\bar\sigma}) \gamma_{k\sigma}^* \dot \gamma_{k\sigma} +  \sum_k (E_0 + E_k - \lambda_k -\frac{2}{\beta} \ln(1+\exp^{-\beta\lambda_k}))+\sum_{k\sigma} \lambda_k \gamma_{k\sigma}^* \gamma_{k\sigma},
\eeq
\end{widetext}
in terms of a set of Bogoliubov quasiparticles,
\beq
\gamma_{k\uparrow}^*&=& +\cos^2 \theta_k c_{k\uparrow}^* + \sin^2 \theta_k c_{-k\downarrow} \\
\gamma_{k\downarrow}&=& -\sin^2 \theta_k c_{k\uparrow}^* + \cos^2
\theta_k c_{-k\downarrow}
\eeq
where $\cos^2\theta_k = \frac{1}{2}(1+\frac{E_k}{\lambda_k})$. Here, $\alpha_k = 2 (\cos k_x
+ \cos k_y )$,  $\eta_k = 2 (\cos k_x - \cos k_y )$, $E_0 =-(2\mu +
\frac{s^2}{U})\varphi^* \varphi$, $E_k =-g_t t \alpha_k -\mu$, $\lambda_k=
\sqrt{E_k^2+\Delta_k^2}$ $\Delta_k= s\varphi^*(1-\frac{2t}{U}\alpha_k)$ and
$g_t=\frac{2\delta}{1+\delta}$ when $\delta=1-n \rightarrow
1-Q+2\varphi^*\varphi$ is a renormalized factor which originates
from the correlated hopping term $(1-n_{i\bar\sigma})
c_{i\sigma}^\dagger c_{j\sigma}(1-n_{j\bar\sigma})$. Starting from
Eq. (\ref{leff}), we integrate over the fermions in Eq. (\ref{geff})
to obtain,

\begin{widetext}
\beq\label{eqG}
G(k,\omega) = \frac{1}{Z}\int [D\varphi^*] [D\varphi] G(k,\omega, \varphi) \exp^{-\sum_k (E_0 + E_k - \lambda_k -\frac{2}{\beta} \ln(1+\exp^{-\beta\lambda_k}) )}
\eeq
\end{widetext}
where
\beq\label{gfinal}
G(k,\omega,\varphi)=\frac{\sin^2\theta_k[\varphi]}{\omega+\lambda_k[\varphi]} + \frac{\cos^2\theta_k[\varphi]}{\omega-\lambda_k[\varphi]}
\eeq
is the exact Green function corresponding to the Lagrangian, Eq. (\ref{leff}).
The two-pole structure of $G(k,\omega,\varphi)$ will figure prominently in the structure of the electron spectral function.
To calculate $G(k,\omega)$, we numerically evaluated the remaining
$\varphi$ integral in Eq. (\ref{eqG}). Since Eq. (\ref{eqG}) is averaged over all values of $\varphi$, we have circumvented the problem inherent in mean-field or saddle-point analyses.  Physically, Eq. (\ref{eqG}) serves to mix (through the integration over $\varphi$) all subspaces with varying number of double occupancies into the low-energy theory.  Hence, it should retain the full physics inherent in the bosonic degree of freedom.

\begin{figure}
\centering
\includegraphics[width=6.cm,angle=270]{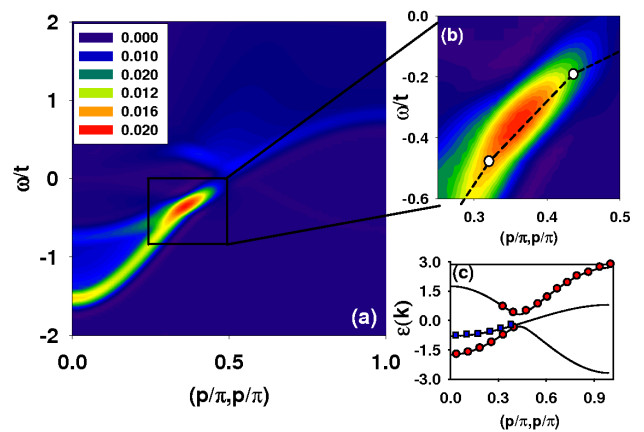}
\caption{(a) Spectral function for filling $n=0.9$ along the nodal direction.  The intensity is indicated by the color scheme.  (b) Location of the low and high energy kinks as inidcated by the change in the slope of the electron dispersion.  (c) The energy bands that give rise to the bifurcation of the electron dispersion.}\label{specf1}
\end{figure}

\begin{figure}
\centering
\includegraphics[width=8.cm]{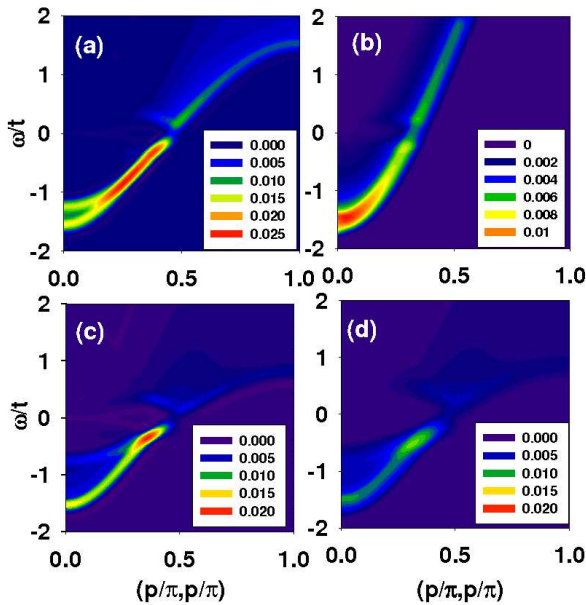}
\caption{Spectral function for two different fillings (a) $n=0.8$ and (b) $n=0.4$ along the nodal direction. The absence of a splitting in the electron dispersion at $n=0.4$ indicates the bifurcation ceases beyond a critical doping. The spectral functions for two different values of the on-site repulsion,
  (c)$U=10t$ and (d)$U=20t$ for $n=0.9$ reveals that the high-energy kink and the splitting of the electron dispersion have at best a weak dependence on $U$.  This indicates that this physics is set by the energy scale $t$ rather than $U$.}\label{specf2}
\end{figure}

The spectral function for $U=10t$ evaluated from Eq. (\ref{gfinal})
and displayed in Figs. (\ref{specf1}) and Fig. (\ref{specf2}) exhibits four key features.  First, regardless of the doping,
there is a low-energy kink in the electron dispersion.  The enlarged region in Fig. (\ref{specf1}a)
shows, in fact, that two kinks exist.  The low-energy kink occurs at roughly $0.2t\approx 100 meV$.  By treating the mass term for the boson as a variable parameter, we verified
that the low-energy kink is determined by the bare mass.  In the effective low-energy theory, the bare mass is $t^2/U$.  This mass is independent of doping.
Experimentally,
the low-energy kink\cite{lenkink} is impervious to doping.  Consequently, the boson offers a natural explanation of this effect that is distinct from the phonon mechanisms which have been proposed\cite{lenkink}.

Second, a high-energy kink appears at roughly $0.5t\approx 250 meV$ which closely resembles the experimental kink at $300meV$\cite{henkink}. At sufficiently high doping (see Figs. (\ref{specf2}a) and (\ref{specf2}b)), the high-energy kink disappears.

Third, experimentally, the high-energy kink is accompanied by a splitting of the electron dispersion into two branches\cite{henkink}.  As is evident, this is precisely the behaviour we find below the chemical potential.
The energy difference between  the two branches achieves a maximum at
$(0,0)$ as is seen experimentally.   A computation of the spectral
function at $U=20t$ and $n=0.9$ reveals that the dispersion as well
the bifurcation still persist.  Further, the magnitude of the
splitting does not change, indicating that the energy scale for the
bifurcation and the maximum energy splitting are set by $t$ and not
$U$. The origin of the two branches is captured in
Fig. (\ref{specf1}c).  The two branches below the chemical potential
correspond to the standard band in the LHB (open square in
Fig. (\ref{specf1}c) on which $\varphi$ vanishes and a branch on which
$\varphi\ne 0$ (open circles in Fig. (\ref{specf1}c).  The two
branches indicate that there are two local maxima in the integrand in
Eq. (\ref{eqG}). One of the maxima, $\varphi=0$, arises from the
extremum of $G(k,\omega,\varphi)$ whereas the other, the effective
free energy (exponent in Eq. (\ref{eqG})) is minimized ($\varphi\ne
0$). Above the chemical potential only one branch survives.  The split
electron dispersion below the chemical potential is consistent with
the composite nature of the electron operator dictated by
Eq. (\ref{cop}).  At low energies, the electron is a linear
superposition of two states, one the standard band in the LHB
described by excitations of the form,
$c_{i\sigma}^\dagger(1-n_{i\bar\sigma})$ and the other a composite
excitation consisting of a bound hole and the charge 2e boson,
$c_{i\bar\sigma}\varphi_i^\dagger$. The former contributes to the
static part of the spectral weight transfer (2x) while the new charge
e excitation gives rise to the dynamical contribution to the spectral
weight transfer. Because the new charge e state is strongly dependent on the
hopping it should disperse as is evident from
Fig. (\ref{specf2})
and also confirmed experimentally.

\begin{figure}
\centering
\includegraphics[width=4.cm, angle=270]{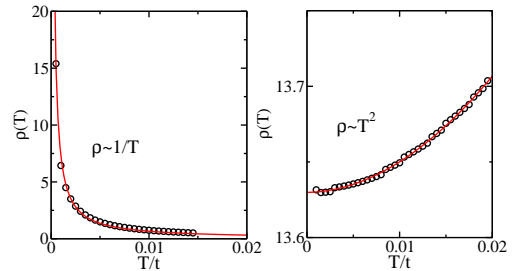}
\caption{(a)dc electrical resistivity as a function of temperature for $n=0.9$ (b) Setting the bosonic degree of freedom to zero kills the divergence of the resistivity as $T\rightarrow 0$. This suggests that it is the strong binding between between the fermionic and bosonic degrees of freedom that ultimately leads to the insulating behaviour in the normal state of a doped Mott insulator.}\label{dcr}
\end{figure}

The formation of the composite excitation,
$c_{i\bar\sigma}\varphi^\dagger$, leads to a pseudogap at the chemical
potential primarily because the charge 2e boson is a local
non-propagating degree of freedom.  The spectral functions for $n=0.9$ and $n=0.8$ both show an absence of spectral weight at the chemical potential.  Non-zero spectral weight resides at the chemical potential
in the heavily overdoped regime, $n=0.4$, consistent with the vanishing of the pseudogap beyond a critical doping away from half-filling. Because the density of states vanishes at the chemical potential, we expect that the electrical resistivity to diverge as $T\rightarrow 0$.  Such a divergence is shown in Fig. (\ref{dcr}a) and is consistent with our previous calculations of the dc resistivity using a local dynamical cluster method\cite{holeloc}.  When the boson is absent (Fig. (\ref{dcr}b)), localization ceases.
Although this calculation does not constitute a proof, it is consistent with
localisation induced by the formation of the bound composite excitation, $c_{i\bar\sigma}\varphi_i^\dagger$.  This state of affairs obtains because the boson is not an inherently dynamical excitation.

\begin{figure}
\centering
\includegraphics[width=4.5cm, angle=270]{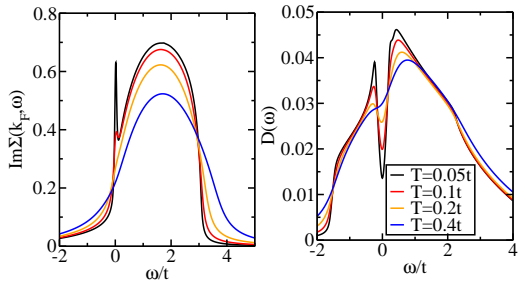}
\caption{The imaginary part of the self energy as the function of
  temperature for $n=0.7$. A peak is developed at $\omega=0$ at low
  temperature which is the signature of the opening of the
  pseudogap. The density of states explicitly showing the pseudogap is
shown in adjacent figure.}\label{self}
\end{figure}

Finally, the imaginary part of the self energy at different temperatures is shown in Fig. (\ref{self}). At low temperature ($T\leq t^2/U$),  the imaginary part of the self-energy at the non-interacting Fermi surface develops a peak at $\omega=0$.  At $T=0$, the peak leads to a divergence.  This is consistent with the opening of a pseudogap. As we have pointed out earlier\cite{zero}, a pseudogap is properly identified by a zero surface (the Luttinger surface) of the single-particle Green function. This zero surface is expected to preserve the Luttinger volume if the pseudogap lacks particle-hole symmetry as shown in the second of the figures in Fig. (\ref{self}).

\subsection{Mid-Infrared Band}

The mid-infrared band (MIB) in the cuprates is a surprise because the optical conductivity in a doped Mott insulator is expected to be non-zero either at the
far-infrared or the ultra-violet or upper-Hubbard-band scale.  While many mechanisms have been proposed\cite{opt3}, no explanation has risen to the fore.
Experimentally, the intensity in the MIB increases with doping at the expense of spectral weight at high energy and the energy scale for the peak in the MIB
is the hopping matrix element $t$.  Since the MIB arises from the
high-energy
 scale, the current theory which accurately integrates
out the high energy degrees of freedom should capture this physics. To obtain a direct link between the conductivity and the spectral function, we work in the non-crossing approximation
\beq\label{cond}
\sigma _{xx} (\omega ) &=& 2
\pi e^2
\int d^2 k\int d\omega '(2t\sin k_x )^2 \nonumber\\
&& \left(  -\frac{f(\omega ')-f(\omega'+\omega)}{\omega} \right)
A(\omega+\omega ',k)A(\omega',k)\nonumber\\
\eeq
to the Kubo formula for the
conductivity where $f(\omega)$ is the Fermi distribution function and $A(\omega,k)$ is the spectral function. At the level of theory constructed here, the
vertex corrections are all due to the interactions with the bosonic degrees of freedom.  Since the boson acquires dynamics only through electron motion and the
leading such term is $O(t^3/U^2)$, the treatment here should suffice to provide the leading behaviour of the optical conductivity.
\begin{figure}
\centering
\includegraphics[width=6.cm]{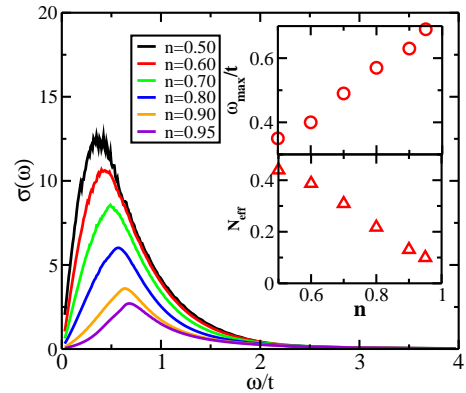}
\caption{Optical conductivity as a function of electron filling, $n$. The peak in the optical conductivity represents the mid-infrared band.  Its origin is mobile double occupancy in the lower-Hubbard band.  The insets show that the energy at which the MIB acquires its maximum value, $\omega_{\rm max}$ is an increasing function of electron filling.  Conversely, the integrated weight of the MIB decreases as the filling increases.  This decrease is compensated with an increased weight at high (upper-Hubbard band) energy scale. }\label{optcond1}
\end{figure}

\begin{figure}
\centering
\includegraphics[width=6.cm]{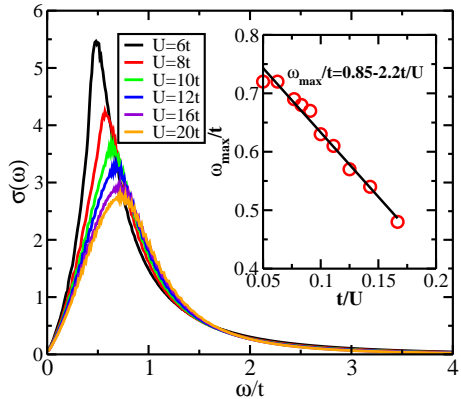}
\caption{Evolution of the optical conductivity for $n=0.9$ as $U$ is varied.
The inset shows the functional form that best describes $\omega_{\rm max}$.  The dominant energy scale is the hopping matrix element $t$ since $t/U$ for the cuprates is $O(1/10)$.}
\label{optcond2}
\end{figure}

 The optical conductivity shown in Fig. (\ref{optcond1}) peaks at $\omega/t\approx .5t$ forming the MIB.  As the inset indicates, $\omega_{\rm max}$ is
 an increasing function of electron filling ($n$) whereas the integrated weight
\beq
N_{\rm eff}=\frac{2m^\ast}{\pi e^2}\int_0^{\Omega_c} d\sigma(\omega)
\eeq
is a decreasing function.  However, $N_{\rm eff}$ does not vanish at half-filling indicating that the mechanism that causes the mid-IR is evident even in the Mott state.   Here we set the integration cutoff to $\Omega_c=2t=1/m^\ast$.
Both the magnitude of $\Omega_{\rm max}$ and its doping dependence as well as the electron filling dependence of the integrated weight are consistent with
that of the mid-infrared band in the optical conductivity in the cuprates\cite{cooper,uchida1,opt1,opt2,opt3}.  To determine what sets the scale for the MIB,
we studied its evolution as a function of $U$.  Figure (\ref{optcond2}) verifies that $\omega_{\rm max}$ is set essentially by the hopping matrix
element $t$ and depends only weakly on $J$. The physical processes that determine this physics are determined by the coupled boson-Fermi terms in the low-energy
theory.  The $\varphi_i^\dagger c_{i\uparrow}c_{i\downarrow}$ term has
a coupling constant of $t$ whereas the $\varphi_i^\dagger b_i$ scales
as $t^2/U$.
Together, both terms give rise to a MIB band that scales as $\omega_{\rm max}/t=0.8-2.21t/U$ (see inset of Fig. (\ref{optcond2})). Since $t/U\approx O(.1)$
for the cuprates, the first term dominates and the MIB is determined
predominantly by the hopping matrix element $t$. Within the
interpretation that $\varphi$ represents a bound state between a
doubly occupied site and a hole, second order perturbation theory with
the $\varphi_i^\dagger b_i$ term mediates the process shown in
Fig. (\ref{midir}).  It is the resonance between these two states that
results in the mid-IR band.  Interestingly, this resonance persists even at half-filling and hence the non-vanishing of $N_{\rm eff}$ at half-filling is not evidence that the cuprates are not doped Mott insulators as has been recently claimed\cite{millis}. In their work, Comanac and colleagues\cite{millis} used
a single-site dynamical mean-field approach.  In such approaches, near-neighbour correlations are absent.

Since the physics in Fig. (\ref{midir}) is not present in projective
models which prohibit double occupancy in the Hubbard basis (not simply
the transformed fermion basis of the t-J model), it is instructive to see what calculations of the optical conductivity in the $t-J$ model
reveal.  All existing calculations\cite{opt2,haulek,prelovsek} on the $t-J$ model find that the MIB scales as $J$.  In some of these calculations, superconductivity
is needed to induce an MIB\cite{haulek} also at an energy scale of $J$.  Experimentally\cite{cooper,uchida1,opt3}, it is clear that the MIB is set by the $t$ scale
rather than $J$.  In fact, since the MIB grows at the expense of spectral weight in the upper-Hubbard band, it is not surprising that the t-J model cannot describe
this physics as first pointed out by
 Uchida, et al.\cite{uchida1}.   The physical mechanism we have
 identified here, Fig. (\ref{midir}) clearly derives from the high energy scale,
 has the correct energy depenence, and hence satisfies the key
 experimental constraints on the origin of the MIB. Since the physics in Fig. (\ref{midir}) is crucial to the mid-IR, it is not surprising that single-site analyses\cite{millis} fail to obtain a non-zero intercept in the extreme Mott limit. The non-zero intercept of $N_{\rm eff}$ is a consequence of Mottness and appears to be seen experimentally in a wide range of cuprates\cite{cooper,int1,int2,int3}

\subsection{Dielectric function: Experimental Prediction}

In the previous section, we have calculated the electronic spectral
function which shows that there are two branches below the chemical
potential. Such physics is explained by the formation of a new
composite excitation, representing a bound state,  consisting of a
bound hole and a charge $2e$ boson, $\varphi_i^\dagger
c_{i\bar\sigma}$.  We demonstrated that for the MIB in the optical
conductivity such an excitation also appears. In principle, these
composite charge excitations should show up in all electric response
functions, for example, the energy loss function, $\Im
1/\epsilon(\omega,\vec q)$, where $\epsilon(\omega,\vec q)$ is the
dielectric function.  We show here that this is the case.

\begin{figure}
\centering
\includegraphics[width=6.cm]{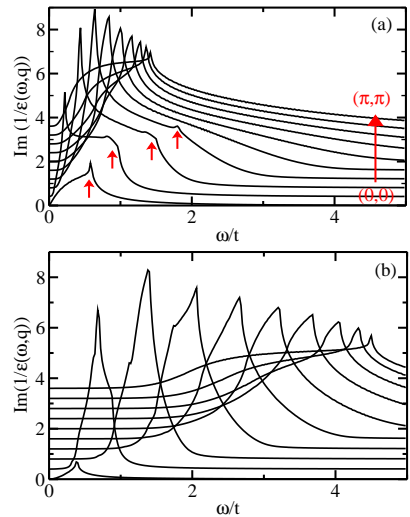}
\caption{The dielectric function, $-\Im 1/\epsilon(\omega,\vec q)$
for $\vec q$ along the diagonal direction is shown for (a) $n=0.9$
and (b) $n=0.6$. Note only the broad feature indicated by the red arrow at $n=0.9$ persists at
$n=0.6$.}
\label{dielect}
\end{figure}

To this end, we calculate the inverse dielectric function,
\beq
\Im \frac{1}{\epsilon(\omega,\vec q)} &=& \pi \frac{U}{v}\sum_p \int d\omega'
(f(\omega')-f(\omega+\omega'))\nonumber\\
&\times& A(\omega+\omega',\vec p+\vec q) A(\omega',\vec p), \eeq
using the non-crossing approximation discussed earlier. Our results
are shown in Fig.(\ref{dielect}) for $n=0.9$ and $n=0.6$ for $\vec
q$ along the diagonal. Two features are distinct. First, there is a
broad band (red arrow in Fig. (\ref{dielect})) with the width of order $t$ that disperses with $\vec q$
for both doping levels. It is simply the particle-hole continuum
which arises from the renormalized bare electron band. The band width is doping
dependent as a result of the renormalization of the band with
doping.  More strikingly,  for $n=0.9$, a sharp peak exists at
$\omega/t\approx .2t$.  It disperses with $q$, terminating when
$\vec q\rightarrow (\pi, \pi)$.   Physically, the sharp peak
represents a quasiparticle excitation of the composite object,
$\varphi_i^\dagger c_{i\bar\sigma}$, the charge 2e boson and a hole.
Therefore, we predict that if this new composite charge excitation,
$\varphi_i^\dagger c_{i\bar\sigma}$, is a real physical entity, as
it seems to be, it will give rise to a sharp peak in addition to the
particle-hole continuum in the inverse dielectric function.  Since
this function has not been measured at present, our work here
represents a prediction.  Electron-energy loss spectroscopy can be
used to measure the inverse dielectric function. Our key prediction
is that momentum-dependent scattering should reveal a sharp peak
that appears at low energy in a doped Mott insulator.  We have
checked numerically the weight under the peaks in the inverse
dielectric function and the sharp peak is important.  Hence, the new
charge $e$ particle we have identified here should be experimentally
observable.

The two dispersing particle-hole features
found here are distinct from a similar feature in stripe models\cite{zaanen}.  In
such models the second branch\cite{zaanen} has vanishing weight and whereas in the
current theory both features are of unit weight.

\subsection{Heat conductivity and heat capacity}

\begin{figure}
\centering
\includegraphics[width=4.cm, angle=270]{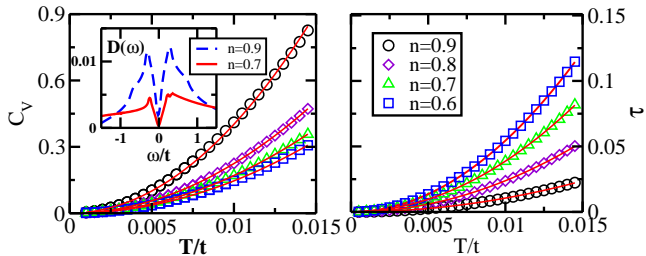}
\caption{(a) Heat capacity, $C_V$, and (b) thermal conductivity, $\tau$, calculated at $n=0.9$.  The solid lines are a fit to $T^2$. Insert: Density of states for $U=10t$ are evaluated at $n=0.9$ and $n=0.7$ respectively.}
\label{thermal}
\end{figure}

As shown by Loram and collaborators\cite{loram}, the heat capacity in the cuprates in the normal state scales as $T^2$.  Quite generally, a density of states that vanishes linearly with energy, that is, a V-shaped gap, yields a heat capacity that scales quadratically with temperature.  The coefficient of the $T^2$ term is determined by the slope of the density of states in the vicinity of the chemical potential. The magnitude of the $T^2$ term should diminish as the doping increases because the slope of the density of states decreases as the pseudogap closes.  As we showed in the previous section, the boson creates a pseudogap.  The energy dependence of the gap is shown in the inset of Fig. (\ref{thermal}). Quite evident is the linear dependence on energy.  The resultant heat capacity shown in Fig.(\ref{thermal}a), calculated via the relationship $C_v=\frac{d \bar E}{dt}$, where the internal energy, $\bar E$, is
\beq
\bar E =\int d\omega D(\omega) \omega f(\omega)
\eeq
and $D(\omega)=\sum_{\vec k}A(\omega, \vec k)$ is the density of states displays a perfectly quadratic temperature dependence in the doping regime where the pseudogap is present as is seen experimentally\cite{loram}.  As it is the boson that underlies the pseudogap, it is the efficient cause of the $T^2$ dependence of the heat capacity.
In our theory, the steeper slope occurs at smaller doping which gives
rise to the largest heat capacity at half filling. This doping
dependence of the heat capacity seems to contradict the experimental
observations\cite{loram}. A key in determining the magnitude of the heat capacity is the spin degrees of freedom.  As we have focused entirely on the bosonic degree of freedom and not on the contribution from the spin-spin interaction terms, we have over-estimated the kinetic energy. Such terms, though they do not affect the pseudogap found here (from Eq. (\ref{cop}) it is clear that the spin-spin terms renormalize the standard fermionic branch in the lower-Hubbard band leaving the new state mediated by $\varphi_i$ untouched), do alter the doping dependence\cite{prelovsek}.

Additionally, the thermal conductivity, $\tau(T)$, can be calculated using the Kubo formula in non-crossing approximation,
\beq
\tau(T)=\frac{e}{4k_B T} \sum_{\vec k}\int \frac{d\omega}{2\pi} (v_{\vec k}^x)^2 \omega^2
\left(-\frac{\partial f(\omega)}{\partial \omega} \right) A(\vec
k,\omega)^2.\nonumber
\eeq
The thermal conductivity shown in Fig.(\ref{thermal}) scales as $T^2$ which is identical to that of the heat capacity.  However, the system exhibits a larger thermal conductivity as the doping increases in contrast to the heat capacity which is decreasing as the doping increases. Physically, this signifies that the carriers are more mobile as the doping increases.


\subsection{T-linear Resistivity}

Over a funnel-shaped region in the $T-x$ plane, the resistivity displays the anomalous linear-T dependence.  The standard explanation\cite{tlin} attributes $T-$ linear resisitivity to quantum criticality. However, one of us has recently shown\cite{tlin} that under three general assumptions, 1) one-paramater scaling, 2) the critical degrees of freedom carry the current and 3) charge is conserved, the resistivity in the quantum critical regime takes the universal form,
\begin{eqnarray}\label{dclimit}
\sigma(\omega=0)=\frac{Q^2}{\hbar}\;\Sigma(0)\;\left(\frac{k_BT}{\hbar
    c}\right)^{(d-2)/z} .
\end{eqnarray}
 Consequently, $T-$linear resistivity obtains (for d=3) only if the dynamical exponent satisfies the unphysical constraint $z<0$. The inability of Eq. (\ref{dclimit}) to lead to a consistent account of
$T-$linear resistivity signifies that 1)
either $T-$linear resistivity is not due to quantum criticality, 2)
additional non-critical degrees of freedom are necessarily the charge
carriers, or 3) perhaps some new theory of quantum criticality can be
constructed in which the single-correlation length hypothesis is relaxed.

\begin{figure}
\centering
\includegraphics[width=7.5cm]{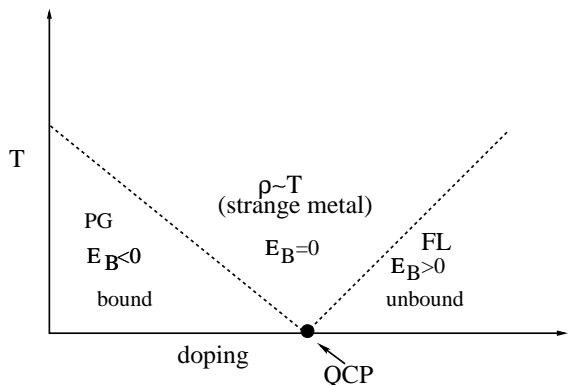}
\caption{Proposed phase diagram for the binding of the holes and bosons that result in the formation of the pseudogap phase.  Once the binding energy vanishes, the energy to excite a boson vanishes.  In the critical regime, the dominant scattering mechanism is still due to the interaction with the boson. T-linear resistivity results anytime $T>\omega_b$, where $\omega_b$ is the energy to excite a boson.  To the right of the quantum critical regime (QCP), the boson is irrelevant and scattering is dominated by electron-electron interactions indicative of a Fermi liquid. The QCP signifies the end of the binding of fermi and bosonic degrees of freedom that result in the pseudogap phase.}
\label{tlin}
\end{figure}

We show that the low-energy theory presented here contains elements
of both (2) and (3) which lead to $T-$linear resistivity.  The
formation of the pseudogap and the divergence of the electrical
resistivity are highly suggestive that a bound state between a hole
and the charge 2e bosonic field, namely the $\varphi^\dagger
c_{i\bar\sigma}$ particle discussed earlier in the electron operator
and the new feature in the dielectric function. Let assume this
state of affairs obtains and the binding energy is $E_B$. As a bound
state, $E_B<0$, where energies are measured relative to the chemical
potential.  Upon increased hole doping, the chemical potential
decreases.  Beyond a critical doping, the chemical potential,
crosses the energy of the bound state.  At the critical value of the
doping where $E_B=0$, the energy to excite a boson vanishes. The
critical region is dominated by electron-boson scattering. In
metals, it is well-known\cite{bass} that above the Debye
temperature, the resistivity arising from electron-phonon scattering
is linear in temperature. In the critical regime, the current
mechanism yields a $T-$linear resisitivity for the same reason.
Namely, in the critical region, the energy to create a boson
vanishes as shown in Fig. (\ref{tlin}) and hence the resistivity
arising from electron-boson scattering should be linear in
temperature. This mechanism is robust as it relies solely on the
vanishing of the boson energy at criticality and not on the form of
the coupling.  To the right of the quantum critical point, standard
electron-electron interactions dominate and Fermi liquid behaviour
obtains.  In this scenario, the quantum critical point coincides
with the termination of the pseudogap phase, or equivalently with
the unbinding of the bosonic degrees of freedom.

\subsection{Final Remarks}

We have shown here that 1) T-linear resistivity, 2) high and low-energy kinks in the electron dispersion, 3) pseudogap phenomena, 4) mid-infrared absorption, 5) $T^2$ heat capacity, and 6) the bifurcation of the electron addition spectrum
all emerge from the bosonic degree of freedom which exists in the
exact low-energy theory of a doped Mott insulator. All the features
found here arise from the charge rather than from the spin dynamics
and hence have nothing to do with the spin-spin interaction of the
$t-J$ model. The essential feature of the charge 2e boson is that it gives the electron substructure, as is observed experimentally.  Further, it does so entirely from a collective mode arising from the strong correlations.  No lattice phenomena need be invoked.  Consequently, we attribute  the anomalies of the normal state of the cuprates to a purely electronic cause which is captured by the physics of the exact-low energy theory of a doped Mott insulator. The key experimental prediction which is sufficient to falsify this theory is the existence of a sharp quasiparticle excitation in the inverse dielectric function. The new quasiparticle represents the collective motion of a double occupancy and a hole as depicted in Fig. (\ref{midir}) which emerges naturally in the exact low-energy theory as a charge $2e$ boson bound to a hole. The existence of such a quasiparticle underscores the fact that double occupancy at low energy requires a hole in its immediate vicinity.  Since this prediction is sharp, experimental falsification of this theory should be straightforward.

\acknowledgements This research was supported in part by the NSF
DMR-0605769
and NSF under Grant No. PHY05-51164 (T.P.C. and P. P.).  We thank
G. A. Sawatzky for insightful discussions and the Kavli Institute of Theoretical Physics for their hospitality where this paper became finalised.

\end{document}